\documentclass[twocolumn,pra,amsmath,amssymb,showpacs]{revtex4}

\usepackage{bm}

\usepackage{graphics}
\usepackage{epsfig}

\def \be{\begin{equation}}
\def \ee{\end{equation}}
\def \bew{\begin{widetext}\begin{equation}}
\def \eew{\end{equation}\end{widetext}}
\def \bmlett{\begin{mathletters}}
\def \emlett{\end{mathletters}}

\def \Wtild{\widetilde{W}}

\def \nbar{\bar{n}_{eq}}

\def \ua{\uparrow}
\def \da{\downarrow}
\def \ra{\rightarrow}

\def \bx{\bar{x}}
\def \bp{\bar{p}}

\def \bs{\bar{s}}

\begin{document}



\title{Using a qubit to measure photon number statistics of a 
driven, thermal oscillator}
\author{A. A. Clerk and D. Wahyu Utami}
\affiliation{ Department of Physics, McGill University, Montr\'{e}al,
 Qu\'{e}bec, Canada, H3A 2T8}
\date{Dec. 9, 2006}

\begin{abstract}
We demonstrate theoretically how photon number statistics of a driven, damped oscillator at finite temperature can be extracted by measuring the dephasing spectrum of a two-level system dispersively coupled to the oscillator; we thus extend the work of Dykman \cite{Dykman87} and Gambetta et al. \cite{Gambetta06}.  We carefully consider the fidelity of this scheme-- to what extent does the measurement reflect the initial number statistics of the mode?  We also derive analytic results for the dephasing of a qubit by a driven, thermal mode, and compare results obtained at different levels of approximation.  Our results have relevance both to experiments in circuit cavity QED, as well as to nano-electromechanical systems.
\end{abstract}
\maketitle

\section{Introduction}

There has been a recent explosion of interest in the properties of quantum oscillators.
On one hand, nano-electromechanical systems (NEMS) (systems comprising of a micron scale oscillator coupled to an electronic conductor) have been used for near quantum-limited position detection, and are thought to be good candidates for the preparation of exotic quantum states and for use in quantum control experiments \cite{Schwab05}.  On the other hand, there has been considerable success in studying circuit QED systems, in which a superconducting qubit is strongly coupled to a single mode in an electromagnetic resonator \cite{Blais04,Wallraff04, Chiorescu04}.  Experiments in these systems have been able to enter the strong coupling regime of cavity QED, achieving such milestones as the resolution of the vacuum Rabi splitting \cite{Wallraff04}.    

A clear goal in both these classes of systems is to reach a regime where the quantum nature of the resonator is unambiguously manifest in measured experimental quantities.   An example of this would be the observation of features associated with the discrete energy states of the oscillator.  This was recently achieved in spectacular fashion in the circuit QED system with a so-called ``number splitting" experiment \cite{Schuster06}.  It was experimentally shown that for a strong dispersive coupling between the qubit and the mode, the qubit dephasing spectrum (i.e.~the time-dependence of the qubit's off-diagonal density matrix element) could be used to probe the photon number distribution in the cavity mode.  Theoretically, this effect was first described by Dykman for the case of a undriven, thermal mode \cite{Dykman87}.  More recently, Gambetta et al.~\cite{Gambetta06} described the effect for the case of a zero-temperature driven mode, a case more relevant to the experiment of Ref.~\onlinecite{Schuster06}.

In the case of NEMS systems, there have been several theoretical proposals for measuring discrete number states, but to date these have been difficult to implement experimentally \cite{Santamore04, Santamore04b, Jacobs06}.  
Given the success of the circuit QED number splitting experiment \cite{Schuster06}, it is natural to ask whether a similar approach could be taken with NEMS systems.  In principle, the answer is yes.  The coupling of nanomechanical oscillators to superconducting qubits has been discussed 
extensively \cite{Armour02, Irish03, Irish05} and is being actively pursued experimentally.
One could thus easily envisage a setup where a 
superconducting qubit is coupled to {\it both} a electromagnetic cavity {\it and} a nanomechanical oscillator \cite{Nori06, Jacobs06}.  The mechanical oscillator would be strongly coupled to the qubit and would generate the number splitting effect, while the electromagnetic cavity would be weakly coupled and would be used simply to read out the qubit dephasing spectrum.  The idea of using the approach of Ref.~\onlinecite{Schuster06} to detect number states in a nano-mechanical oscillator is highly attractive.  From an experimental perspective, it is not as demanding as previous proposals (e.g. strong non-linearities in the oscillator are not required).  In addition, we will see that such an experiment offers the possibility of providing an {\it unambiguous} signature of the discrete number states of the oscillator; no extensive post-experiment interpretation is required.

The above discussion notwithstanding, there remain unresolved theoretical issues with the number splitting effect that should be addressed before contemplating a NEMS version of the experiment.  These issues are also relevant to the circuit-QED experiment of Ref.~\onlinecite{Schuster06}.  First, as NEMS oscillators do not typically have frequencies much larger than temperature, one should now consider {\it both} the effect of finite temperature and a driving force; this regime is not addressed by existing theory \cite{Dykman87, Gambetta06}.  In this paper, we calculate analytically the qubit dephasing spectrum for a qubit dispersively coupled to a driven, finite temperature oscillator, using a conceptually simple Wigner function approach.  We recover the previous results in the appropriate limits.  Our results also provide analytic expression for the long-time oscillator-induced dephasing rate of the qubit in the driven, thermal case (c.f Eqs.~(\ref{eq:ThermDeph}) and (\ref{eq:DriveDeph})).  Note that the dephasing rate in the absence of a drive was considered recently by Serban et al. in Ref.~\onlinecite{Serban06}.

A second important issue we address here is the question of measurement fidelity: to what extent is the qubit dephasing spectrum a {\it faithful} measurement of the mode's number statistics?  It was already pointed out in Refs.~\onlinecite{Gambetta06, Schuster06} that for a generic driving force, there was not a simple connection between the observed peaks in the qubit dephasing spectrum and the photon number distribution in the mode.  We demonstrate that in general, back-action effects arise from the finite damping of the oscillator; these imply that the dephasing spectrum of the qubit will fail to perfectly reflect the initial photon distribution of the oscillator.  Nonetheless, we demonstrate that for a sufficiently strong oscillator-qubit coupling, one can 
reliably extract the initial photon distribution of the oscillator. 

\section{Heuristic discussion of the number splitting effect}

We consider a qubit which is dispersively coupled to a harmonic oscillator.  Ignoring for the moment the damping and driving of the oscillator, the Hamiltonian of the system is given by:
\begin{eqnarray}
		H	& = &	
			H_{0} + H_{int}		\\ 
		H_{0} & = & 
			\frac{\hbar \Omega_{qb}}{2} \sigma_z + 
			\hbar \Omega \left( a^{\dagger} a + \frac{1}{2} \right) 
			\\
		H_{int} & = & 
			\lambda \hbar \Omega \left( a^{\dagger} a + \frac{1}{2} \right) \sigma_z 
			\label{eq:HInt}
\end{eqnarray}
Here, $\hbar \Omega_{qb}$ is the energy splitting of the qubit, $\Omega$ is the oscillator frequency, and $\lambda$ is a dimensionless measure of the interaction strength.
As discussed, e.g.~ in Ref.~\onlinecite{Blais04}, an interaction of this type can arise from a Jaynes-Cumming type coupling in the limit where the frequency difference $| \Omega_{qb} - \Omega |$ is large (i.e.~one transforms the starting Hamiltonian to eliminate the first-order qubit-mode interaction; the dispersive interaction above then results at second order in the original coupling strength).  It has recently been suggested \cite{Serban06} that it is important to keep non-rotating terms in the interaction Hamiltonian $H_{int}$ (i.e.~terms proportional to $a^2$ and $(a^{\dagger})^2$).  {\it We find that such terms play  no role in the weak damping, weak coupling limit of interest} (i.e.~$\lambda \ll 1, \gamma / \Omega \ll 1$); we discuss this further in what follows.  Note that in this limit, the effects of the qubit-oscillator interaction may still be significant, as $\lambda \Omega / \gamma$ can be large.

The quantity we wish to study is the qubit's off-diagonal density matrix element.  Letting $|\ua \rangle, | \da \rangle$ denote $\sigma_z$ eigenstates and $\hat{\rho}$ the density matrix of the oscillator-qubit system, we define:
\begin{eqnarray}
	\hat{\rho}_{\ua \da}(t) & = &   
			\rho_{\ua \da}(t) 
			\equiv \langle \ua | \hat{\rho}(t) | \da \rangle 
		\\
	\rho_{\ua \da}(t) & = & 
		\textrm{Tr}_{osc} \phantom{\cdot} \hat{\rho}_{\ua \da}(t)
\end{eqnarray}
Note that $\rho_{\ua \da}(t)$ is indeed a measurable quantity; one could imagine extracting it using state tomography, as has recently been demonstrated for superconducting qubits \cite{Martinis06a}.  An alternate method, as discussed in Refs.~\cite{Dykman87,Gambetta06} and used in the experiment of 
Ref.~\onlinecite{Schuster06}, is to measure $\rho_{\ua \da}(t)$ via an absorption experiment.  In such a scheme, one attempts to excite the qubit from its ground state via a time-dependent field at frequency $\omega_{rf}$ which couples to $\sigma_x$.  Within Fermi's golden rule and the quantum regression theorem \cite{ScullyBook97}, the absorption rate $\Gamma_{abs}(\omega_{rf})$ is given by:
\begin{eqnarray}
	\Gamma_{abs}(\omega_{rf}) & = &
		|A|^2 \int_{-\infty}^{\infty} dt e^{+i \omega_{rf} t} \left \langle 
			\sigma_-(t) \sigma_+(0) 
		\right \rangle 
	\nonumber \\		
		& = & 2 |A|^2 \textrm{Re } 
		\left[ \int_{0}^{\infty} dt e^{+i \omega_{rf} t}
		\rho_{\ua \da}(t) \right]
		\label{eq:AbsorptionSpectrum}
\end{eqnarray}
Here, the second line assumes the quantum regression theorem, and $A$ is a matrix element proportional to the strength of the driving field.

\subsection{Number splitting interpretation of $\rho_{\ua \da}(t)$}

The simplest expectation for the behaviour of  $\rho_{\ua \da}(t)$ follows from the observation that a) this quantity should oscillate at the qubit splitting frequency, and b) the dispersive coupling in Eq.~(\ref{eq:HInt}) implies qubit splitting frequency depends on the number of quanta in the mode.  This leads to the expected form:
\begin{eqnarray}
	\rho_{\ua \da}(t) = e^{-i \Omega_{qb} t} \sum_{n=0}^{\infty} P(n) e^{-2 i \lambda \Omega
	 \left(n + \frac{1}{2} \right) t}
	\rho_{\ua \da}(0)
	\label{eq:RhoTInterp}
\end{eqnarray}
where $P(n)$ is the number distribution of the harmonic mode.  Eq.~(\ref{eq:RhoTInterp}) would of course be exact in the absence of damping or forcing of the oscillator.  It tells us that the {\it time-dependence} of $\rho_{\ua \da}$ {\it at a fixed coupling strength} contains information about the oscillator photon number distribution at $t=0$:  if one Fourier-transforms $\rho_{\ua \da}(t)$, one expects a set of  discrete peaks with a spacing $\Delta \omega = 2 \lambda \Omega$, with the weight of each peak yielding the probability $P(n)$ of the corresponding number state.  

In the more general case where we have to contend with damping and forcing of the oscillator, Eq.~(\ref{eq:RhoTInterp}) is no longer a priori valid.  The interaction $H_{int}$ now fails to commute with the total oscillator Hamiltonian, and thus there will be a back-action associated with the measurement.  Heuristically, turning on the coupling to the qubit can now itself cause a change in the number statistics of the oscillator.  For example, the coupling to the qubit causes the oscillator's frequency to change; at non-zero damping, this will cause it to re-equilibrate with the bath and thus change its number distribution.  
We can nonetheless attempt to use Eq.~(\ref{eq:RhoTInterp}) to extract the number statistics of the oscillator (i.e.~use the weights of peaks in the spectrum of $\rho_{\ua \da}$ to extract the  distribution of number states in the oscillator).  We will thus refer to Eq.~(\ref{eq:RhoTInterp}) as the ``number splitting" interpretation of $\rho_{\ua \da}$; in what follows, we will assess the validity of such an interpretation.   

\subsection{Full counting statistics interpretation of $\rho_{\ua \da}(\lambda)$}

A second, somewhat different way to connect $\rho_{\ua \da}(t)$ to number statistics is to use the {\it coupling} dependence of this quantity (as opposed to its time dependence) to extract the oscillator number statistics.  This is the procedure used in defining the full counting statistics of transmitted charge in mesoscopic electron conductors \cite{Levitov03}.  The application of this approach to the present system will be discussed in a forthcoming work \cite{ClerkWahyuInPrep}.

\section{Equation of motion for $\rho_{\ua \da}$}

We now wish to calculate $\rho_{\ua \da}(t)$ in the presence of both oscillator damping and driving.  The Hamiltonian of our system $H$ thus takes the form:
\begin{eqnarray}
	H & = & 
		H_{0} + H_{int} 
	- \hbar \Omega F(t) \left(a + a^{\dagger} \right)
			\\ 
	&&
		  + \left(a + a^{\dagger} \right) 
	\sum_{j} v_j \left(b_j + b_j^{\dagger} \right) + 
	\sum_j \hbar \omega_j \left(b^{\dagger}_j b_j + \frac{1}{2} \right)
	\nonumber 
\end{eqnarray}
$F(t)$ is the external drive on the oscillator.  The $b_j$ modes correspond to a thermal bath responsible for the intrinsic dissipation of the oscillator; we choose the density of states of these modes and the  couplings $v_j$ to yield a standard Ohmic dissipation.  Note we are considering an ideal case where the qubit has no intrinsic (i.e.~oscillator-independent) relaxation or dephasing.  Such intrinsic processes can be accounted for by multiplying $\rho_{\ua \da}(t)$ for the ideal case by an extra decaying factor 
$\exp \left( -t \left(\frac{1}{T_2} + \frac{1}{2 T_1} \right) \right)$ \cite{Gambetta06}.

We are interested in the evolution of $\hat{\rho}_{\ua \da}(t)$, which is an operator in the oscillator's Hilbert space.  We are also interested in the weak damping limit $\gamma \ll \Omega$ which is relevant to experiment.  In this limit, one can start with an exact path-integral description and then rigorously derive a local-in-time equation of motion for $\rho_{\ua \da}$; this is in complete analogy to the derivation of the usual weak-coupling Caledeira-Leggett master equation \cite{Caldeira89}.  This equation can also be rigorously justified in the high temperature $k_B T \gg \hbar \Omega$ limit.
Setting $\Omega_{qb}=0$ (as it plays no role in what follows), one has:
\bew
	\frac{ \partial}{\partial t} \hat \rho_{\ua \da}  = 
		-\frac{i}{\hbar} 
			[ H_{osc}, \hat{\rho}_{\ua \da} ]
			+ i \Omega \frac{ F(t) } {\Delta x_0} 
				[ x, \hat{\rho}_{\ua \da}  ] 
			- \frac{i \gamma}{2 \hbar}
			[ x, \{ p, \hat{\rho}_{\ua \da} \} ]
			- \frac{D}{\hbar^2} \left [ x, [ x, \hat{\rho}_{\ua \da} ] \right ]
			- \frac{i}{\hbar} \lambda \{ H_{osc}, \hat{\rho}_{\ua \da} \}
		\label{eq:CLMaster}
\eew
Here, $H_{osc} = p^2/(2m) + m \Omega^2 x^2 / 2$, $\gamma$ is the damping rate of the oscillator due to the bath, and ($\beta = 1 / (k_B T)$)
\begin{eqnarray}
D = m \gamma \frac{ \hbar \Omega}{2} \coth(\beta \hbar \Omega / 2)
\end{eqnarray}
is the corresponding momentum diffusion constant.  We also use $\Delta x_0$ and $\Delta p_0$ to denote the zero-point variances of the oscillator.  Note that unlike the standard quantum optics master equation 
(used in Refs.~\onlinecite{Dykman87, Gambetta06}), the above equation has counter-rotating terms (i.e.~proportional to $a^2$ and $(a^{\dagger})^2$) and is 
hence not in Linblad form.  We will comment more on this difference in what follows. 

To proceed, we will represent $\rho_{\ua \da}$ by its Wigner function:
\begin{eqnarray}
	W(x,p;t) \equiv
		\frac{1}{2 \pi \hbar} \int dy e^{-i p y/ \hbar}
		\rho_{\ua \da}(x+\frac{y}{2},x-\frac{y}{2};t)
\end{eqnarray}
Writing Eq.~(\ref{eq:CLMaster}) in terms of the Wigner function, one finds
\begin{widetext}
\begin{eqnarray}
	\partial_t W(x,p;t) & = & 
	\left[ -\frac{p}{m} \partial_x + 
	\left( m \Omega^2 x - \frac{\hbar \Omega}{\Delta x_0} F(t) \right) \partial_p 
	+ \gamma \partial_p \cdot p + D \partial^2_p  
	- i \frac{\lambda}{\hbar} \left[
		\frac{p^2}{m} + m \Omega^2 x^2 - 
		\frac{\hbar^2}{4} \left(
			\frac{1}{m} \partial^2_x +
			m \Omega^2 \partial^2_p \right) \right] \right] W(x,p;t)
				\nonumber \\
				\label{eq:NewFP}
\end{eqnarray}
\end{widetext}
The $\lambda$-independent terms above are the usual terms in a classical Fokker-Planck equation for a damped, driven oscillator, while the terms proportional to 
$\lambda$ yield distinctly non-classical contributions.  As the additional quantum terms are at most quadratic in $x,p$ and their derivatives, we may solve this equation in the same way one would solve the classical Fokker-Planck equation: one makes a Gaussian ansatz.  Note that a similar approach was used recently by Serban et al.~\cite{Serban06} to study the dephasing of a qubit by a thermal, damped mode.

Thus, we write:
 \begin{eqnarray}
 	W(x,p) = \int \frac{dk}{2 \pi} \int \frac{dq}{2 \pi} e^{i k x} e^{i q p}
		\Wtild(k,q)
\end{eqnarray}
and make the Gaussian ansatz:
\begin{eqnarray}
	\frac{\Wtild(k,q;t)}{\Wtild(0,0;0)} & = & 
		e^{i \nu}
		e^{i k \Delta x_0 \cdot \bx + i q \Delta p_0 \cdot \bp }
		\times  \\ 
	&&	\exp \Big[
			-\frac{1}{2} \big(
				k^2 (\Delta x_0)^2 \sigma_{x} + 
				q^2 (\Delta p_0)^2 \sigma_{p} 
				\nonumber \\
	&&	+ 
				2 k q (\Delta x_0) (\Delta p_0) \sigma_{x p} 
			\big) 
		\Big]. \nonumber
\end{eqnarray}
The Gaussian is parameterized by six time-dependent parameters:  the means $\bx(t)$, $\bp(t)$, the variances $\sigma_x(t)$ and $\sigma_p(t)$, the cross-correlation $\sigma_{xp}(t)$ and the overall phase $\nu(t)$; note that we have scaled all quantities by the zero-point variances of the oscillator.   Substituting the above ansatz into the 
appropriately transformed version of Eq.~(\ref{eq:CLMaster}), we find that it is satisfies the equation as long as the Gaussian parameters satisfy the following equations (we set $\Omega = 1$ from this point onwards):
\begin{eqnarray}
	\frac{d \bx}{dt}  & = &
		\bp - i \lambda 
			\left( \sigma_x \bx + \sigma_{xp} \bp \right) 
		\label{eq:BXEqn} \\
	\frac{d\bp }{dt}  & = &
		- \bx + 2 F(t) - \gamma \bp  
		- i \lambda  \left( \sigma_p \bp + \sigma_{xp} \bx \right)
		\label{eq:BPEqn} \\
	\frac{ d \sigma_x}{dt} & = &
		+ 2 \sigma_{xp} 
		- i \lambda \left( \sigma_x^2 + \sigma_{xp}^2 - 1 \right) 
		\label{eq:SigmaXEq} \\
	\frac{ d \sigma_p}{dt} & = &
		- 2 \sigma_{xp} - 2 \gamma (\sigma_p - \tilde{T} ) - 
		i \lambda \left( \sigma_p^2 + \sigma_{xp}^2 - 1 \right) 
		\label{eq:SigmaPEq} \\
	\frac{ d \sigma_{xp}}{dt} & = &
		\sigma_{p} - \sigma_{x} - \gamma \sigma_{xp}
		- i \lambda \sigma_{xp} \left( \sigma_x + \sigma_p \right) 
		\label{eq:SigmaXPEq}\\
	\frac{ d \nu}{dt} & = & - \frac{\lambda}{2} \left( \sigma_x + \sigma_p \right)
		- \frac{\lambda}{2} \left( \bx^2 + \bp^2 \right) \label{eq:NuEqn}
\end{eqnarray}		
Here, we have defined:
\begin{eqnarray}
	\tilde{T} \equiv \coth{ \beta \hbar \Omega / 2} = 2 \nbar + 1 
\end{eqnarray}

The above equations have a simple form.  The $\lambda$-independent terms describe the usual dynamics of a driven dissipative oscillator, and have a completely classical form.  The $\lambda$-dependent terms act as quantum source terms, and cause the means and variances of the Gaussian to acquire imaginary parts.  
From Eq.~(\ref{eq:NuEqn}), we find that the decay of $\rho_{\ua \da}$ is determined from these quantities:
\begin{eqnarray}
	 \frac{\rho_{\ua \da}(t)}{\rho_{\ua \da}(0)}   & = & 
	 \frac{\Wtild(k=0,q=0;t)}{\Wtild(k=0,q=0;0)} \nonumber \\
		& = & \Lambda_1(t) \cdot \Lambda_2(t)
			\label{eq:Suppress}
\end{eqnarray}
with
\begin{eqnarray}
	\Lambda_1(t) & = & 
		\exp \left[
			-i \frac{\lambda}{2}
			\int_0^t dt'  \left(
			\sigma_x(t') + \sigma_p(t') \right) 
			\right ]
	\label{eq:Lambda1}			\\
	\Lambda_2(t) & = &
		\exp \left[
			-i \frac{\lambda}{2} 
			\int_0^t dt' \left(
			\bx^2(t') + \bp^2(t') \right)
			\right]
	\label{eq:Lambda2}			
\end{eqnarray}

The two factors above correspond to two distinct contributions to the decay of $\rho_{\ua \da}$, and hence to the number splitting effect.  The factor $\Lambda_1(t)$ describes a {\it purely} thermal effect:  from Eqs. (\ref{eq:SigmaXEq}) and (\ref{eq:SigmaPEq}), we see that $\sigma_x$ and $\sigma_p$ acquire negative imaginary parts, causing $\Lambda_1(t)$ to decay from its initial value of $1$.  This effect only occurs at finite temperature, as at zero temperature $\sigma_x(0) = \sigma_y(0) = \tilde{T} = 1$ and are hence time independent.  Note that $\Lambda_1(t)$ is completely independent of the means of the Gaussian (i.e.~$\bx$ and $\bp$) and hence of the driving force $F(t)$.  

In contrast, the factor $\Lambda_2(t)$ in Eq.~(\ref{eq:Suppress}) describes the effect of a finite drive on $\rho_{\ua \da}$:  if the means $\bx$ and $\bp$ are non-zero, Eq.~(\ref{eq:BXEqn})-(\ref{eq:BPEqn}) indicate that they will acquire an imaginary part and cause this factor to decay.

\section{Dephasing spectrum in the purely thermal case}

We first consider the case of zero drive (i.e.~$F(t)=0$), and take the oscillator to initially be in a thermal state: $\bx=\bp=0$, $\sigma_x = \sigma_p = \tilde{T}$, and $\sigma_{xp} = 0$.  Note that this corresponds to an oscillator in equilibrium with frequency $\Omega$; one might argue that for the absorption experiment, the initial condition should correspond to an equilibrium oscillator with frequency $\Omega - \lambda$.  This difference is easy to include; in the weak coupling limit of interest ($\lambda \ll 1$), the corrections are lower-order in $\lambda$ and hence not-significant. 

As there is no drive, Eqs.~(\ref{eq:BXEqn}) and (\ref{eq:BPEqn}) imply $\bx=0$, $\bp=0$ for all time, and thus the factor $\Lambda_2(t) = 1$.  The evolution of $\rho_{\ua \da}$ is the completely determined by the factor $\Lambda_1(t)$.  To proceed with the analysis,  it is convenient to write:
\begin{eqnarray}
\sigma_x & = & 1 + \bs + \Delta s \\
\sigma_p & = & 1 + \bs - \Delta s
\end{eqnarray}  
The equations for the Gaussian parameters thus become:
\begin{eqnarray}
	\frac{d \nu}{dt} & = & \lambda(1 + \bs) 
		\label{eq:NuEqn2}\\  
	\frac{d \bs}{dt} & = &
		- i \lambda \left[
			2 \bs + \bs^2 + (\Delta s)^2 + \sigma_{xp}^2 
			\right]
			\nonumber \\
	&&
			-  \gamma(\bs - \Delta s)  + 2 \gamma \nbar 
			\label{eq:BsEqn} \\
	\frac{d \Delta s}{dt} & = &
		- 2 i \lambda (1 + \bs)  (\Delta s)  
			\nonumber \\
	&&
		+ \gamma(\bs - \Delta s)  - 2 \gamma \nbar 
		+ 2 \sigma_{xp} \\
	\frac{d \sigma_{xp}}{d t} & = &
		- \gamma \sigma_{xp} - 2 \Delta s
		- 2 i \lambda \sigma_{xp} \left[1 + \bs \right]
		\label{eq:XPEqn}
\end{eqnarray}		

As we start in a thermal state, both $\Delta s(0)=0 $ and $\sigma_{xp}(0)=0$; further, if we took the limit $\gamma \ra 0$, these quantities would remain zero for all time.
Thus, in the weak damping ($\gamma \ll 1$) limit of interest, it will be an excellent approximation to set $\Delta s = \sigma_{xp}=0$ in Eq.~(\ref{eq:BsEqn}); this results in the approximate equation:
\begin{eqnarray}
	\frac{d \bs}{dt} & = &
		- i \lambda \left[
			2 \bs + \bs^2  
			\right]
			-  \gamma \left( \bs  - 2 \nbar \right) 
			\label{eq:BsEqn2} 
\end{eqnarray}
Within this approximation,  $\Lambda_1(t)$ is completely determined by Eq.~(\ref{eq:BsEqn2}) and Eq.~(\ref{eq:NuEqn2}).  Note crucially that our approximation does not amount to doing perturbation theory in $\lambda$ or $\gamma$, as terms to all orders in the ratio $\lambda / \gamma$ are retained.  Also note that if one had started by 
using the quantum optics version of Eq.~(\ref{eq:CLMaster}) (i.e.~drop all $a^2$ and $(a^{\dagger})^2$ terms), one would
immediately find Eq.~(\ref{eq:BsEqn2}) without any additional approximation.  We thus see that the differences between the Linblad and non-Linblad forms of the master equation play no role in the weak coupling, small damping limit of interest.  In what follows, we will explicitly discuss the correction terms that arise when one uses the full equation Eq.~(\ref{eq:BsEqn}) versus the approximate version Eq.~(\ref{eq:BsEqn2}).

\subsection{Results for dephasing spectrum}

We can now solve directly Eq.~(\ref{eq:BsEqn2}) for $\bs$.  We find:
\begin{eqnarray}
	(\sigma_x + \sigma_p)/2 & = & 
		\frac{i \gamma}{2 \lambda} + \alpha
		\frac{1 + M e^{-2 i \lambda \alpha t} }    
		{1 - M e^{-2 i \lambda \alpha t} } 
		\label{eq:sbarsoln}
\end{eqnarray}
where
\begin{eqnarray}
	\alpha & = & 
		\sqrt{ \left(1 - \frac{i \gamma}{2 \lambda} \right)^2 
		- \frac{2 i \gamma \nbar }{\lambda}  } 
	\label{eq:AlphaEqn} \\
	M & = &
		\frac{
			2 e^{-\beta \Omega} + \left( e^{-\beta \Omega}-1 \right)
				\left( \alpha - \left( 1 -  \frac{ i \gamma}{2 \lambda} \right) \right)
			}
			{
			2 + \left( e^{-\beta \Omega}-1 \right)
				\left( 1 - \alpha + \frac{ i \gamma}{2 \lambda} \right)
			}		
		\label{eq:MEqn}
\end{eqnarray}
Note that at $T=0$, we have simply $\alpha = 1 - i \gamma / (2 \lambda)$ and $M=0$.

Combining Eqs.~(\ref{eq:Lambda1}) and (\ref{eq:sbarsoln}) we find:
\begin{eqnarray}
	\frac{\rho_{\ua \da}(t)}{\rho_{\ua \da}(0)} & = & 
		\Lambda_1(t) \nonumber \\
	& = & 
		e^{ \gamma t / 2} e^{-i \lambda \alpha t}
		\frac{1 - M}{1 - M e^{-2 i \lambda \alpha t} } \\
		& = &
			e^{ \gamma t /2} (1-M)
			\sum_{n=0}^{\infty}
				M^n e^{-i (2n + 1) \lambda \alpha  t}
				\label{eq:ThermalRho}
\end{eqnarray}
Note that the result above coincides exactly with Eq.~(11) in Ref.~\onlinecite{Dykman87}.  Unlike that work, we did not start by making a rotating-wave approximation in the bath-oscillator interaction, but instead took a weak-damping limit.

Consider first the $\gamma \ra 0$ limit.  In this limit, one finds $M = e^{-\beta \Omega}$ and $\alpha = 1$.  Eq.~(\ref{eq:ThermalRho}) thus yields $\rho_{\ua \da}(t)$ in the form of Eq.~(\ref{eq:RhoTInterp}) with a thermal number distribution.  As expected, $\rho_{\ua \da}$ is a completely faithful measure of the oscillator number statistics in this limit, as there is no back-action:  at zero damping, the mode has no way to re-equilibrate its number distribution in response to a change in its frequency.

In the case of small but finite $\gamma$, we see that $|M|$ will play the role of the Boltzman weight $e^{- \beta \omega}$, and that $\alpha$ acts to renormalize the coupling $\lambda$.  For small damping, the Fourier spectrum of $\rho_{\ua \da}(t)$ will again consist of a set of equally-spaced discrete peaks occurring at frequencies $\omega_n$.  The peaks will now have a Lorentzian form,  each having a non-zero  width $\gamma_n$.  To leading order in $\gamma$ we have:
\begin{eqnarray}
	\omega_n & = &
		(2 n + 1) \lambda \left[
			1 + \frac{1}{2}  
			\left( \frac{\gamma}{\lambda} \right)^2
			\nbar \left( 1 + \nbar \right)
			\right]	\\
		\gamma_n & = &
			\gamma \left[ \nbar + n (1 + 2 \nbar ) \right]
\end{eqnarray}
Further, to order $\gamma^2$, the weighting factor $|M|$ is given by:
\begin{eqnarray}
	|M| \simeq
		e^{-\beta \Omega} \left[ 1 -
			\frac{\gamma^2}{4 \lambda^2} \left( 1 + 2 \nbar \right) \right]
\end{eqnarray}
We can use $|M|$ to define an effective temperature $T_{eff}$ via 
$|M| \equiv e^{-\hbar \Omega / (k_b T_{eff})}$.  This is the temperature one would associate with the oscillator based on using Eq.~(\ref{eq:RhoTInterp}) to interpret the peaks in the spectrum of $\rho_{\ua \da}(t)$.  To leading order in $\gamma / \lambda$, we have:
\begin{eqnarray}
	T_{eff} & = & T \times
		\left(1 - \frac{\gamma^2}{4 \lambda^2} \frac{k_B T}{\hbar \Omega}
			(1 + 2 \nbar) \right) 
	\label{eq:Teff}
\end{eqnarray}

Thus, in the weak-but finite damping limit, the probability distribution extracted from the time dependence of $\rho_{\ua \da}$ at fixed coupling strength $\lambda \gg \gamma$ (i.e.~using Eq.~(\ref{eq:RhoTInterp})) is very close to the initial, equilibrium photon number distribution of the oscillator.  Note that there is a one-to-one correspondence between the peaks in the spectrum and the peaks expected from Eq.~(\ref{eq:RhoTInterp}).  The main effect of the damping is to give the peaks associated with each number state a finite width in frequency.  There is also deviation between the weight of each peak and a pure thermal weight; the finite damping makes the oscillator appear somewhat {\it colder} than it should.  On a heuristic level, it is not difficult to understand why we have a correspondence.  The time-scale for back-action effects to set in is $\sim 1/\gamma$-- this is how long the oscillator takes to re-equilibrate its number distribution given the change in frequency brought on by the coupling to the qubit.  For $\lambda \gg \gamma$, the qubit effectively ``measures" the number statistics well before they have a chance to be changed by the back-action.  

\subsection{Results for oscillator-induced dephasing rate}

It is also interesting to consider the limiting behaviour of 
$\rho_{\ua \da}(t)$ in the long-time limit, $t \gg 1/\gamma$.  We can define the long-time dephasing rate $\Gamma_{\varphi,th}$ and frequency shift 
$\Delta_{th}$ of the qubit via:
\begin{eqnarray}
	\Delta_{th}	& \equiv &  -\textrm{Im } \lim_{t \ra \infty} 
		\frac{\log \rho_{\ua \da}(t)}{t} \\
	\Gamma_{\varphi,th} & \equiv & 
		 - \textrm{Re } \lim_{t \ra \infty} 
		\frac{\log \rho_{\ua \da}(t)}{t} 
\end{eqnarray}
This yields:
\begin{eqnarray}
	\Delta_{th}	
		& = & 
		\textrm{Re} \lambda \alpha 
		\nonumber \\
		& = & 
			\frac{\gamma}{2}
				\textrm{Im }
					\left[
						\sqrt{ \left(
								1 + \frac{2 i \lambda}{\gamma} \right)^2
						+ \frac{8 i \lambda}{\gamma} \nbar 
						}				  
			\right] 
			\label{eq:ThermalDOmega} \\
	\Gamma_{\varphi,th} & = & 
		 -\frac{\gamma}{2} - \textrm{Im} \lambda \alpha 
		 \nonumber \\
		& = &
			\frac{\gamma}{2}
				\textrm{Re }
					\left[
						\sqrt{ \left(
								1 + \frac{2 i \lambda}{\gamma} \right)^2
						+ \frac{8 i \lambda}{\gamma} \nbar 
						}
				 - 1 
			\right]  \nonumber \\
	 \label{eq:ThermDeph}
\end{eqnarray}
We find that the thermal oscillator-induced dephasing rate may be written as $\gamma$ times a function which depends only on the ratio $\lambda / \gamma $ and the temperature; in typical experimental situations, one has both $\gamma \ll 1$ and $\lambda \ll 1$, but the ratio $\lambda / \gamma$ can be arbitrary.  
Note also that the thermal dephasing rate of the qubit vanishes at zero temperature.  Expanding in $\gamma / \lambda$ we have:
\begin{eqnarray}
	\Delta_{th} & = &  \lambda +
		\frac{\gamma^2}{2 \lambda} \nbar (1 + \nbar)
		+ O(\gamma^4) \\
	\Gamma_{\varphi,th} & = & 
			\gamma \nbar - 
				\frac{\gamma^3}{4 \lambda^2}
				\nbar \left(1 + \nbar \right) \left(1 + 2 \nbar \right)
				+ O\left( \frac{\gamma^5}{\lambda^4} \right)
				\nonumber \\
		 	\label{eq:ThermDeph2}
\end{eqnarray}
In contrast, in the high-temperature limit $\nbar \gg 1$, we find for the dephasing rate:
\begin{eqnarray}
	\Gamma_{\varphi,th} & \rightarrow &
		\sqrt{ \gamma \lambda \nbar}
		\label{eq:ThermDephHighT}
\end{eqnarray}
Finally, expanding $\Gamma_{\varphi,th}$ as a function of the coupling yields the usual Fermi's golden rule result:
\begin{eqnarray}
	\Gamma_{\varphi,th} \rightarrow
		\lambda^2 \frac{4 \nbar (1 + \nbar)}{\gamma}
		= \lambda^2 S_{n}(\omega=0)
	\label{eq:ThermDephSmallLam}
\end{eqnarray}
where $S_n(\omega=0)$ is the zero-frequency charge noise of the oscillator at zero coupling (i.e.~$\lambda = 0$).  Note that the high-temperature limit of the Golden rule result and the full answer are not the same!  This can be understood from the fact that higher order terms in $\lambda$ become increasingly important at higher temperatures;
the Golden rule result is only valid if $\lambda \nbar / \gamma \ll 1$ is satisfied.

Eqs.~(\ref{eq:ThermDeph})-(\ref{eq:ThermDephHighT}) for the thermal dephasing rate are valid within the small-damping approximation made in Eq.~(\ref{eq:BsEqn2}), or equivalently, within a rotating-wave description of the bath-oscillator coupling.  One can also find $\Gamma_{\varphi,th}$ {\it without} making these approximations by solving the full Eqs.~(\ref{eq:NuEqn2})-(\ref{eq:XPEqn}) for a stationary $\bs$.  Results are given in the Appendix.  One again finds that the thermal dephasing rate vanishes at zero temperature, and has the same high-temperature limit as 
Eq.~(\ref{eq:ThermDephHighT}).  One also finds that corrections to the result 
Eq.~(\ref{eq:ThermDeph}) are higher-order in $\gamma$.  If we keep the ratio $\gamma/\lambda$ fixed and expand in $\gamma$, we find:
\begin{eqnarray}
	\left[ \Gamma_{\varphi,th} \right]_{full}
	= \Gamma_{\varphi,th} + O(\gamma^3)
\end{eqnarray}
This confirms our prior claim that the error introduced by the approximation of Eq.~(\ref{eq:BsEqn2}) is small in the weak damping $\gamma \ll 1$ limit.

Finally, we have also calculated the thermal dephasing rate for a qubit-oscillator interaction Hamiltonian which contains counter-rotating wave terms, i.e.~:
\begin{eqnarray}
	H_{int,2} = \frac{\lambda \hbar \Omega}{2} \left(a + a^{\dagger} \right)^2
	\label{eq:Hint2}
\end{eqnarray}
The thermal dephasing rate of the qubit for the interaction Hamiltonian $H_{int,2}$ was previously considered by Serban et al. \cite{Serban06}.
Using a completely analogous procedure as the above, we can find the oscillator induced dephasing rate of the qubit with this modified interaction; results are provided in the appendix.  As expected, we find that corrections to the result of Eq.~(\ref{eq:ThermDeph}) are negligible in the $\gamma, \lambda \ll 1$ limit of interest.


\section{Dephasing spectrum in the presence of an oscillator drive}

We now consider the additional effect of a driving force $F(t)$ on the oscillator.  The thermal factor $\Lambda_1(t)$ considered in the previous section will be unchanged by the presence of the force, assuming the initial state of the oscillator has variances consistent with being in equilibrium.  However, because of the drive, the means of the Gaussian (i.e.~ $\bx$ and $\bp$) will be time-dependent; as a result, the factor $\Lambda_2(t)$ (c.f.~Eqn.~(\ref{eq:Lambda2})) will also contribute to the suppression of $\rho_{\ua \da}(t)$ and to the number splitting effect.  To characterize this, we must calculate $\bx(t)$ and $\bp(t)$.

We begin by writing:
\begin{eqnarray}
	a_1(t) & = & \frac{1}{2}(\bx(t) + i \bp(t)) \\
	a_2(t) & = & \frac{1}{2}(\bx(t) - i \bp(t) ) 
\end{eqnarray}
Note that in general,  $a_1(t) \neq [a_2(t)]^*$ as $\bx$ and $\bp$ have complex parts;
we only have $a_1(t) = [a_2(t)]^*$ at $t=0$, when both $\bx$ and $\bp$ are real.  
Also note that with our rescaling, $a_1$ and $a_2$ correspond to the usual $a$ for a harmonic oscillator.  As $\bx^2 + \bp^2 = 4 a_2 a_1$, we have:
\begin{eqnarray}
	\Lambda_2(t) = 
		\exp \left(- 2 i \lambda \int_0^t dt' a_2(t') a_1(t') \right)
		\label{eq:DriveContrib}
\end{eqnarray}
Note the similarity to Eq.~(5.16) in Ref.~\onlinecite{Gambetta06}.  In the positive P-function approach used in that work, it was important to have a different coherent states associated with the two states of the qubit ($| \ua \rangle, | \da \rangle$).  In our approach, this ``doubling" of degrees of freedom manifests itself in the fact that $\bar{x}$ and $\bar{p}$ have both real and imaginary parts (i.e.~ $a_2 \neq (a_1)^*$).

 We use Eqs.~(\ref{eq:BXEqn}) and (\ref{eq:BPEqn}) to obtain equations for $a_1$ and $a_2$.  Assuming the drive is near the oscillator frequency $\Omega$, and that $\lambda \ll 1$, it is safe to make a rotating wave approximation to these equations.  Writing $F(t) = \textrm{Re } \left[ F e^{-i \omega_d t} \right ]$, we thus find (recall we have set $\Omega = 1$):
\begin{eqnarray}
	\frac{d}{dt} a_1 & = &
		\left[
			-i - \frac{\gamma}{2} 
				- i \lambda \left( \frac{ \sigma_x + \sigma_p}{2} \right)
		\right] a_1  + \frac{i}{2} F_{+}(t)
			  \\  	
	\frac{d}{dt} a_2 & = &
		\left[
		i - \frac{\gamma}{2}
			- i \lambda  \left( \frac{ \sigma_x + \sigma_p}{2} \right)
		\right]
			a_2  - \frac{i}{2} \left[ F_{+}(t) \right]^*
\end{eqnarray} 
where $F_{+}(t) = F e^{-i (\omega_d - 1) t}$.
We see that due to the coupling $\lambda$, the average variance $\sigma_x(t) + \sigma_p(t)$ acts as a time-dependent modulation of the oscillator frequency.  This average variance is again completely determined from 
Eq.~(\ref{eq:BsEqn}) of the previous section, and is completely independent of the force $F(t)$ or the initial value of $a(t)$.  In addition,  as we have seen, it will have an imaginary part.

The above equations are easily solved to yield:
\begin{eqnarray}
	a_1(t) & = & g_1(t,0) a_1(0) +
		\frac{i}{2} \int_0^t dt' g_1(t,t')  F_+(t') 
		\label{eq:a1soln} \\  
	a_2(t) & = & g_2(t,0) a_2(0) -
		\frac{i}{2} \int_0^t dt' g_2(t,t')  \left[ F_+(t') \right]^* 
		\label{eq:a2soln}
\end{eqnarray}
with
\begin{eqnarray}
	g_{1}(t,t') & = &
		e^{-i  (t-t')} e^{- \gamma (t-t')/2}
		\times \\
		&&
		\exp\left[
			- i \lambda 
			\int^t_{t'} dt'' 
				\left( 
					\frac{\sigma_x(t'') + \sigma_p(t'')}{2}  
				\right)
			\right]   \nonumber \\
	g_2(t,t') & = &
		e^{i  (t-t')} e^{- \gamma (t-t')/2}
		\times  \label{eq:g2Defn} \\
		&&
		\exp\left[
			 - i \lambda 
			\int^t_{t'} dt'' 
				\left(
					\frac{\sigma_x(t'') + \sigma_p(t'')}{2}  
				\right)
			\right]  \nonumber
\end{eqnarray}

Eqs.~(\ref{eq:DriveContrib}), (\ref{eq:a1soln}) -(\ref{eq:g2Defn}) are enough to completely determine the drive contribution $\Lambda_2(t)$ to the decay of $\rho_{\ua \da}(t)$.  In the limit of zero temperature, there is no thermal contribution to the decay of $\rho_{\ua \da}$ (i.e.~ there is no zero-temperature dephasing), and $\sigma_x = \sigma_p = 1$ for all times.  This corresponds to the limit considered
in Ref.~\onlinecite{Gambetta06}; as we will see, the current approach yields the same answer found in that work.  In the more general case of a non-zero temperature, the thermal number splitting will also influence the ``drive" contribution to the number splitting, through the dynamics of $\sigma_x(t) + \sigma_p(t)$.

To proceed, we make use of the small-$\gamma$ solution for $\bs(t)$ obtained in the previous section
(c.f. Eq.~(\ref{eq:sbarsoln})).  We stress again that this result would be exact {\it if} one started by making a rotating wave approximation in Eq.~(\ref{eq:CLMaster}), as is commonly done in the quantum optics community.
We also stress that corrections to this approximation are negligible in the weak-damping, weak coupling ($\lambda \ll 1, \gamma \ll \Omega$) limit of interest.
We thus find:
\begin{eqnarray}
	g_1(t_1,t_2) & = &
		e^{-i (1 + \lambda \alpha) (t_1-t_2)}
		\frac{1 - M e^{-2 i \lambda \alpha t_2} }
		{1 - M e^{-2 i \lambda \alpha t_1} }   \\
	g_2(t_1,t_2) & = &
		e^{i (1 - \lambda \alpha) (t_1-t_2)}
		\frac{1 - M e^{-2 i \lambda \alpha t_2} }
		{1 - M e^{-2 i \lambda \alpha t_1} }  
\end{eqnarray}
The parameters $M$ and $\alpha$ are temperature dependent, and are given by 
Eqs.~(\ref{eq:AlphaEqn}) and (\ref{eq:MEqn}) respectively.  Note that $g_1$ and $g_2$ are {\it not} time-translational invariant, as they are sensitive to the initial time $t=0$ at which the coupling was turned on.  Also note that in the zero temperature limit, we have $M \ra 0$ and $\lambda \alpha \ra \lambda - i \gamma /2$.  The propagators $g_1$ and $g_2$ then reduce to the expected propagators for a damped oscillator having a frequency of $\Omega \pm \lambda$.  The propagators also have this form at long times ($t_1 \gamma, t_2 \gamma \gg 1$) in the finite-temperature case.  

We can now write $a_1(t)$ and $a_2(t)$ as:
\begin{eqnarray}
	a_1(t) & = & a_{1,s} \cdot e^{-i \omega_d t} + \tilde{a}_1(t)\\
	a_2(t) & = & a_{2,s} \cdot e^{i \omega_d t} + \tilde{a}_2(t) 
\end{eqnarray}
Here, $\tilde{a}_1(t)$ and $\tilde{a}_2(t)$ describe the transient behaviour of $a_1$ and $a_2$.  This decomposition then allows us to write the drive-dependent contribution to $\rho_{\ua \da}(t)$, $\Lambda_2(t)$, as:
\begin{eqnarray}
	\Lambda_2(t) 
		& \equiv & \exp \left[
		-i \Delta_{dr}t - \Gamma_{\varphi, dr} t \right] \times
		\exp \left[ Q_{trans}(t) \right]
		\nonumber
\end{eqnarray}
where
\begin{eqnarray}
	-i \Delta_{dr} - \Gamma_{\varphi, dr}  
	& \equiv & 
		-2 i \lambda a_{1,s} a_{2,s}  \\
	Q_{trans}(t) & \equiv &	
			-2 i \lambda \int_0^t dt' \Big[
				 \tilde{a}_1(t') \tilde{a}_2(t')
		 \label{eq:QDefn} \\
			&&
				 + a_{1,s} \tilde{a}_2(t') e^{-i \omega_d t'}
				 + a_{2,s} \tilde{a}_1(t') e^{+i \omega_d t'}
				\Big]
				\nonumber	
\end{eqnarray}
We thus see that the long time behaviour of $a_1, a_2$ will set the drive contribution $\Gamma_{\varphi,dr}$ to the dephasing rate of the qubit (i.e.~the exponential decay of $\rho_{\ua \da}$ at long times), while the transient behaviour of these two functions will be (presumably) responsible for the number splitting effect.

\subsection{Results for the drive contribution to the dephasing rate}

In the presence of a drive, the total dephasing rate of the qubit (i.e.~the exponential decay rate of $\rho_{\ua \da}(t)$ at long times) will be given by:
\begin{eqnarray}
	\Gamma_{\varphi} = \Gamma_{\varphi,th} + \Gamma_{\varphi,dr}
\end{eqnarray}
The purely thermal contribution $\Gamma_{\varphi,th}$ is given in Eq.~(\ref{eq:ThermDeph}).  Using the results of this section, we can calculate the drive contribution to the dephasing rate.  One first finds the stationary amplitudes $a_{1,s}$ and $a_{2,s}$.  Letting $\delta \equiv \Omega - \omega_d$, we have:
\begin{eqnarray}
	a_{1,s} & = & \frac{i F / 2}{i \delta + i \lambda \alpha} 
		= 
		\frac{i F / 2}
		{i (\delta + \Delta_{th}) + 
			\frac{\gamma}{2} + \Gamma_{\varphi,th}  }\\
	a_{2,s} & = & \frac{-i F^* / 2}{-i \delta + i \lambda \alpha} 
		= 
		\frac{-i F^* / 2}
		{-i (\delta - \Delta_{th}) + 
		\frac{\gamma}{2} + \Gamma_{\varphi,th} }
\end{eqnarray}
Here, $\Delta_{th}$ is the thermal frequency shift of the qubit (c.f. Eq.~(\ref{eq:ThermalDOmega})); to lowest order in $\gamma$, $\Delta_{th} = \lambda$.  Note that the stationary amplitudes are temperature-dependent!

Using these expressions, and letting $\gamma_{th} \equiv
( \gamma + 2 \Gamma_{\varphi,th})$, we find that the drive-dependent contribution to the qubit dephasing rate takes the form:
\begin{eqnarray}
	\Gamma_{\varphi,dr} & = & \frac{|F|^2}{2} 
		\frac{\lambda \Delta_{th} \gamma_{th} }
		{ \left[
			(\delta + \Delta_{th})^2 + 
				\left(\frac{\gamma_{th}}{2}\right)^2
		\right]
		\left[
			(\delta - \Delta_{th})^2 + 
				\left( \frac{\gamma_{th}}{2} \right)^2
		\right]
		}
		\nonumber \\
		\label{eq:DriveDeph}
\end{eqnarray}
Note that at zero temperature, $\Delta_{th} = \lambda$ and $\Gamma_{\varphi,th}=0$.  In this limit, Eq.~(\ref{eq:DriveDeph}) reduces to the expression 
found in Ref.~\onlinecite{Gambetta06}.  We see that the effect of finite temperatures is to 
both provide an extra term in the dephasing rate of the qubit (i.e.~ $\Gamma_{\varphi,th}$ in Eq.~(\ref{eq:ThermDeph})), and also to modify the form of the drive-dependent contribution to the dephasing rate.


\subsection{Results for the drive contribution to the dephasing spectrum}

As discussed, the transient parts of $a_1(t)$ and $a_2(t)$ will lead to oscillating contributions to $\rho_{\ua \da}$.  Using the above results, one finds that these transients are given by:
\begin{eqnarray}
	\tilde{a}_1(t) & = &
		\frac{ 1}{1 - M e^{-2i \lambda \alpha t}}
			\left[
				A_1 \cdot e^{-i(1 + \lambda \alpha)t}
				+ B_1 e^{- i (\omega_d + 2 \lambda \alpha ) t}
			\right] 
			\nonumber \\
			\\
	\tilde{a}_2(t) & = &
		\frac{ 1}{1 - M e^{-2i \lambda \alpha t}}
			\left[
				A_2 \cdot e^{i(1 - \lambda \alpha)t}
				+ B_2 e^{i (\omega_d - 2 \lambda \alpha ) t}
			\right] 
			\nonumber \\
\end{eqnarray}
with
\begin{eqnarray}
	A_1 & = &
		(1-M) \times
				\label{eq:A1Eqn}  \\
		&&
			 \left[
			  a(0) - \frac{i F / 2}{i \delta + i \lambda \alpha}
			\left(
				1 -  \frac{2 M}{(1-M)} 
					\frac{i \lambda \alpha}
						{i \delta - i \lambda \alpha}
			\right)
		\right] 
	\nonumber \\
	A_2 & = &
		(1-M) \times 
		 \label{eq:A2Eqn} \\
		&&
			\left[
			  \left[ a(0) \right]^* - \frac{-i F^* /2 }{- i \delta + i \lambda \alpha}
			\left(
				1 + \frac{2 M}{(1-M)} 
										\frac{i \lambda \alpha}
						{i \delta + i \lambda \alpha}
			\right)
		\right] 
	\nonumber  \\
	B_1 & = &
		- 2 i \lambda \alpha M \frac{i F / 2}
			{ \left(i \delta + i \lambda \alpha \right)
			\left( i \delta - i \lambda \alpha \right) }	\\
	B_2 & = &
		- 2 i \lambda \alpha M \frac{-i F^* / 2}
			{ \left(i \delta + i \lambda \alpha \right)
			\left( i \delta - i \lambda \alpha \right) }
\end{eqnarray}

The contribution to the number splitting from the transients of $a_1$ and $a_2$ is given by Eq.~(\ref{eq:QDefn}).  In general, the result will be a complex pattern of peaks, with no simple relation to the original photon number distribution of the mode.  There are however several important situations where things simplify, as we now show.  

\subsubsection{Zero temperature limit}

In the zero temperature limit, we have $B_1 = B_2 = 0$, and:
\begin{eqnarray}
	A_1 & = & a(0) - a_{1,s} \\
	A_2 & = & [a(0)]^* - a_{2,s} 
\end{eqnarray}
Only three terms are present in $Q_{trans}$:
\begin{eqnarray}
	&& \exp\left[ Q_{trans}(t) \right] = 
	\exp \left[
		\frac{A_1 A_2}{1 - i \frac{\gamma}{2 \lambda} }
		\left( e^{-2 i \lambda \alpha t} - 1 \right) 
	\right] \times
	\nonumber \\
	&& 
	\exp \left[
		 \frac{ 2 a_{1,s} A_2 \lambda}{\lambda \alpha - \delta}
		\left( e^{i \delta t} e^{-i \lambda \alpha t} - 1 \right)
	\right] \times \nonumber \\
	&&
	\exp \left[
		 \frac{ 2 a_{2,s} A_1 \lambda}{\lambda \alpha + \delta}
		\left( e^{-i \delta t} e^{-i \lambda \alpha t} - 1 \right)	
	\right]
		\label{eq:QDriveOnly}
\end{eqnarray}

Each factor above would yield a Poisson distribution of oscillating exponentials.  The combination of all three would thus yield a complex distribution of peaks in the spectrum of $\rho_{\ua \da}$, with weights being given as a convolution of three Poisson distributions.  Ref.~\onlinecite{Gambetta06} considered an initial condition where the oscillator is in a state relaxed to the qubit being down: $a(0) = (a_{2,s})^*$.  This is the correct initial state for an absorption experiment 
(c.f. Eq.~(\ref{eq:AbsorptionSpectrum})), where the coupling is always on, and the qubit is initially in its ground state.  In this case $A_2 = 0$, and only the third factor in Eq.~(\ref{eq:QDriveOnly}) contributes.  One thus finds that Fourier transform of $\rho_{\ua \da}(t)$ consists of a set of Poisson-distributed peaks whose frequency spacing is approximately (in the small-$\gamma$ limit) $\Delta \omega = \delta + \lambda$.  {\it Note that if we assumed the spectrum of $\rho_{\ua \da}$ directly reflected the number statistics of the oscillator (c.f. Eq.~(\ref{eq:RhoTInterp})), we would expect a spacing between peaks of $\Delta \omega = 2 \lambda$}.  Also note that though one does obtain a Poisson distribution of peaks, the mean of this distribution does not correspond to $|a(0)|^2$ (i.e.~the mean of the original number distribution before the measurement was turned on).  {\it It is unclear to what extent we can say we have ``measured" the initial number distribution of the oscillator in this case}.  
Note that if we specialize further, and chose the detuning of the driving force to be $\delta = \lambda$, then the simple expectation of Eq.~(\ref{eq:RhoTInterp}) {\it does} hold \cite{Gambetta06}: one obtains peaks with the expected spacing $2 \lambda$, and whose weights correspond to a Poisson distribution with mean $|a(0)|^2$ .

Let us now consider a different approach.  Returning to the full expression, we see that only the first factor in Eq.~(\ref{eq:QDriveOnly}) yields a set of peaks with the expected spacing of $\Delta \omega = 2 \lambda$, regardless of the detuning $\delta$.  Consider the ``strong measurement regime", where $\lambda$ 
is much larger than both $\delta$ and $\gamma/2$.  Further,  take the initial state of the oscillator to correspond to $\lambda = 0$ (i.e.~a driven damped oscillator, but no coupling to the qubit).  One then has $|a(0)| \gg |a_{1,s}|, |a_{2,s}|$.  In this limit, one can approximate $A_1 \simeq a(0)$, $A_2 \simeq a(0)^*$.  Only the first 
factor in Eq.~(\ref{eq:QDriveOnly}) will contribute to the peak spacing; the second and third terms will be suppressed by the small factors $|a_{1,s} / a(0)|$ and $|a_{2,s}/a(0)|$ respectively.  Thus:
\begin{eqnarray}
	\exp\left[ Q_{trans}(t) \right] \simeq 
		\exp\left[
			\frac{|a(0)|^2}{1 - \frac{i \gamma}{2 \lambda}}
			\left(
				e^{-2 i \lambda t}e^{-\gamma t} - 1
			\right)
		\right]
\end{eqnarray}

The Fourier transform of $\rho_{\ua \da}(t)$ will thus consist of a Poisson-distributed set of peaks with the correct spacing of $\Delta \omega = 2 \lambda$; moreover, in the small damping limit $\gamma \ll \lambda$, the mean of this distribution agrees with the mean of the original state,  $|a_0|^2$.  We have thus shown that at zero temperature, one can 
reliably extract the initial number distribution of the driven oscillator (arbitrary detuning $\delta$) from the time-dependence of $\rho_{\ua \da}$ using the ``number splitting" interpretation of Eq.~(\ref{eq:RhoTInterp}).  To do so, one must keep the qubit-oscillator coupling $\lambda$ off before $t=0$; further, for $t>0$, the coupling must be suitably large:  $\lambda \gg \delta, \gamma$.  Note that one may still have $\lambda \ll 1$.  
It is worth emphasizing that the initial condition we consider here (i.e.~$H_{int}=0$ before $t=0$) is not appropriate for the absorption experiment of Ref.~\onlinecite{Schuster06}.  However, it could conceivably be realized in experiments 
involving a nano-mechanical oscillator capacitively coupled to a qubit, as in such systems the qubit-oscillator coupling is proportional to a tunable voltage.

\subsubsection{Drive number splitting, finite temperature}

We now consider the transient drive contribution in the presence of a finite temperature.  We again consider the ``strong measurement limit" $\lambda \gg \delta, \gamma$.  In this limit, the initial amplitude of the oscillator $|a(0)|$ is much larger than $|a_{1,s}|$ or $|a_{2,s}|$.  Thus,  the only significant term in $Q_{trans}(t)$ is the one proportional to $A_1 A_2$, as this is the only term which scales as $|a(0)|^2$.  Using $A_1 A_2 \simeq (1-M)^2 |a(0)|^2$
(c.f. Eqs.(\ref{eq:A1Eqn})-(\ref{eq:A2Eqn})),  we find:
\begin{eqnarray}
	Q_{trans} & = &
		-2 i \lambda A_1 A_2 \int_0^t dt' \frac{ e^{-2 i \lambda \alpha t'}}
			{ \left(1 - M e^{-2 i \lambda \alpha t'} \right)^2 }	\\
		& = &
		\frac{ |a(0)|^2 }{\alpha} 
			\left( e^{-2 i \lambda \alpha t}  -1 \right)
			\left[1 + M
				\frac{e^{-2 i \lambda \alpha t} -1 }{ 1 - M e^{-2 i \lambda \alpha t'}  }  
			\right] 		
			\nonumber \\					
\end{eqnarray}		

If we now combine the above results for the drive contribution to $\rho_{\ua \da}(t)$ with the thermal contribution $\Lambda_1(t)$ (c.f. Eq.~(\ref{eq:ThermalRho})) we find:
\begin{eqnarray}
	\rho_{\ua \da}(t)  
	& = &
		e^{-i \lambda \alpha t}
		e^{-i \Delta _{dr} t} e^{-\Gamma_{\varphi,dr} t}
		\times
		\nonumber \\
	&&
		\tilde{P}(k = -2 \lambda \alpha;T_{eff},F_{eff},\delta)		
	\\
	& = &
		e^{-i \Delta_{dr} t} e^{-\Gamma_{\varphi,dr} t}
		\times 
		\nonumber \\
	&&
		\sum_n P(n;T_{eff},F_{eff},\delta) 
		e^{- 2 i \lambda \alpha \left( n + \frac{1}{2} \right) t} 
	\label{eq:FinalDrivenRho}
\end{eqnarray}
Here, $P(n;T,F,\delta)$ is the photon number distribution of an oscillator at temperature $T$ which is driven by a force of magnitude $F$ and detuning $\delta$;  
$\tilde{P}(k) = \sum_n e^{i k} P(n)$ is the corresponding generating function.  The explicit form of this distribution is given in the Appendix.  To see the correspondence most directly, make the substitutions $e^{-\beta \Omega} \ra M$ and $k
\ra -2 \lambda \alpha t$ in the generating function given in Eqs.~(\ref{eq:GenFunc1}) - (\ref{eq:GenFunc3}).  We find that in the small damping limit, $T_{eff}$ and $F_{eff}$ 
appearing in Eq.~(\ref{eq:FinalDrivenRho}) are almost identical to the initial temperature and driving force on the oscillator: $T_{eff}$ is again given by Eq.~(\ref{eq:Teff}) and
\begin{eqnarray}
	F_{eff} & = & \frac{F}{\sqrt{|\alpha|}}
		\nonumber \\
	& \simeq &
		F\left(1 - \frac{\gamma^2}{8 \lambda^2}
			\left( 1 + 8 \nbar + 8 \nbar^2 \right)
		\right)
\end{eqnarray}

Note the close connection between Eq.~(\ref{eq:FinalDrivenRho}) and the naive
``number splitting" expectation for $\rho_{\ua \da}(t)$ given in Eq.~(\ref{eq:RhoTInterp}).  
{\it We have thus shown that in the ``strong measurement limit" $\lambda \gg \gamma, \delta$, the spectrum of $\rho_{\ua \da}(t)$ accurately reflects the number statistics of the mode}:  in this limit, each peak in the spectrum of $\rho_{\ua \da}$ corresponds to a number state in the mode, and the corresponding peak weight to $P(n)$.  The width $\gamma_n$ of the peak corresponding the photon number $n$ is easily found to be:
\begin{eqnarray}
	\gamma_n & = &
		\Gamma_{\varphi,dr} +
			\left(2n+1 \right)
			\left( \Gamma_{\varphi,th} + \frac{\gamma}{2} \right) - \frac{\gamma}{2}
\end{eqnarray}

In Fig. 1, we show results for the dephasing spectrum $S(\omega)$ for various values of the qubit - oscillator coupling strength $\lambda$, where:
\begin{eqnarray}
	S(\omega) & = &
		\left|
			\int_0^{\infty} dt e^{i \omega t}
			\left(
				\frac{\rho_{\ua \da}(t)}{\rho_{\ua \da}(0)}
				e^{i \lambda t}
			\right) \right|^2
		\label{eq:SDefn}
\end{eqnarray}
Note we have shifted the spectrum so that the peaks associated with number states should appear at frequencies $\omega$ equal to even multiples of $2 \lambda$.  One also expects from Eq.~(\ref{eq:RhoTInterp}) that the weights of the different peaks peaks in the spectrum should correspond to the oscillator number distribution.  These weights may be approximately obtained from $S(\omega)$ via:
\begin{eqnarray}
	P(m)_{approx} = 
		\frac{ 1 }{ \pi \sqrt{h_m}}  \int_{(2m-1)\lambda}^{(2m+1) \lambda}
		d \omega S(\omega)
	\label{eq:PmApprox}
\end{eqnarray}
where $h_m$ is the maximum of $S(\omega)$ in the interval $\left[
(2m-1) \lambda, (2m+1)\lambda) \right]$.  Number distributions obtained 
from $S(\omega)$ using this procedure are shown in Fig. 2.  Note that one could be more sophisticated in extracting the weights of peaks from $S(\omega)$; 
we use Eq.~(\ref{eq:PmApprox}) only for simplicity.  

\begin{figure}[tbh]
\begin{center}
\includegraphics[width=1.00 \columnwidth]{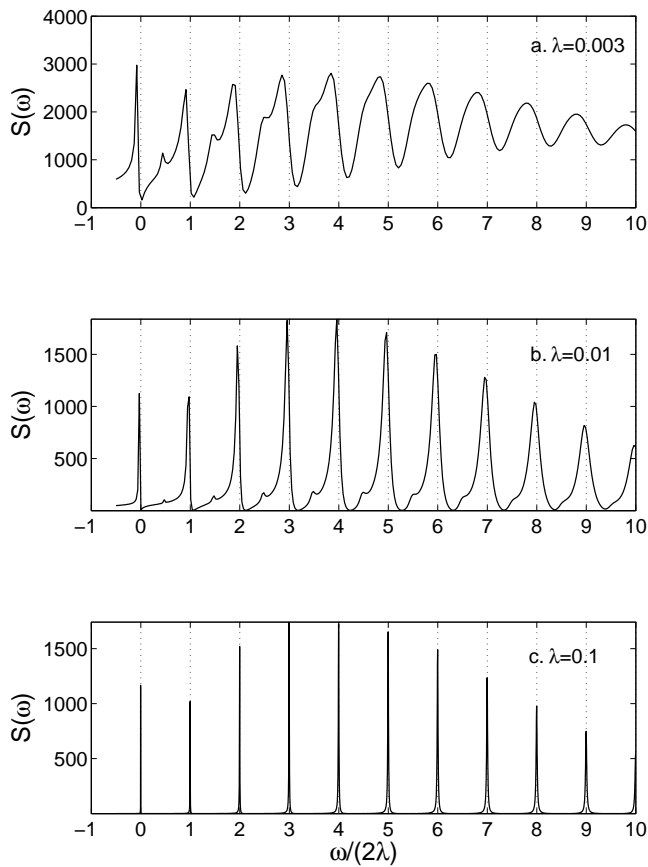}
\caption{Fourier transform of the dephasing spectrum, 
$S(\omega)$ (c.f. Eq.~(\ref{eq:SDefn})) for different values of $\lambda$.  
For each plot, we have taken the oscillator damping $\gamma = 10^{-4}$, 
a driving force of magnitude $|F| = 3 \gamma$ and detuning $\delta=0$, a temperature which corresponds to $\nbar = 1$, and an initial  
condition which corresponds to the driven thermal oscillator at zero qubit-oscillator coupling.  Based on the number splitting interpretation of Eq.~(\ref{eq:RhoTInterp}), one expects to see peaks in $S(\omega)$ whenever $\omega = 2 \lambda$.  For small values of $\lambda$, $S(\omega)$ is in poor correspondence to the number distribution; note that there are additional peaks occurring roughly at $\omega$ equal to odd
multiples of $\lambda$.  These unwanted peaks become increasingly insignificant as $\lambda$ is increased.  The weights of the peaks in plot c) are in good agreement with the expected number state distribution for the initial oscillator state (see Fig. 2).   } 
\label{fig:SpectrumPlot}
\end{center}
\end{figure}

\begin{figure}[tbh]
\begin{center}
\includegraphics[width=1.00 \columnwidth]{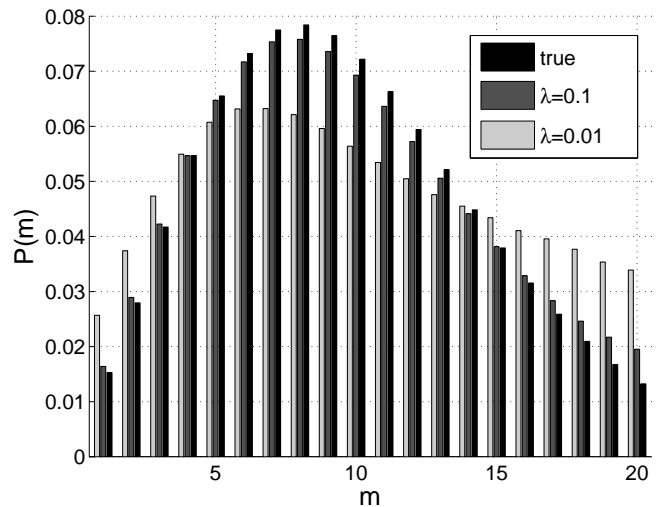}
\caption{Number distributions derived from the dephasing spectrum, using the same oscillator parameters as in Fig.~1.  The black bars correspond to the actual photon number distribution associated with the initial oscillator state.  The remaining bars correspond to distributions obtained from the dephasing spectrum via Eq.~(\ref{eq:PmApprox}).  For $\lambda=0.01$, there is a good correspondence.  Note that the deviations from the true distribution in this case are mainly due to the inaccuracy of using Eq.~(\ref{eq:PmApprox}) to obtain the weight of peaks.}
\label{fig:SpectrumPlot2}
\end{center}
\end{figure}

\section{Conclusions}

In this paper, we have used a conceptually simple Wigner function approach to study the physics of a qubit dispersively coupled to a driven, thermal harmonic mode.  In general, back-action effects prevent the qubit's dephasing spectrum (i.e.~the Fourier spectrum of $\rho_{\ua \da}(t)$) from being a perfect mirror of the number statistics of the initial oscillator state.  We have shown however that for a sufficiently strong qubit-oscillator coupling, one can indeed reliably extract the photon distribution of the driven, thermal mode from the dephasing spectrum.  We have also calculated the oscillator-induced dephasing of the qubit in the case of a thermal, driven mode.  

We thank Alexandre Blais and Mark Dykman for useful conversations.  This work was supported by NSERC, FQRNT and the Canadian Institute for Advanced Research.

\begin{appendix}

\section{Corrections to the thermal dephasing rate}

In the main text, the thermal dephasing rate is calculated starting from the master equation Eq.~(\ref{eq:CLMaster}), using a weak-damping approximation.  As discussed, this approximation is equivalent to having dropped terms proportional to $a^2$ and $(a^{\dagger})^2$ in Eq.~(\ref{eq:CLMaster}), as is often done in the quantum optics literature; this approximation was also used in Refs.~\onlinecite{Dykman87, Gambetta06}.  In this section, we discuss corrections arising when one goes beyond this approximation.

\subsection{Results from the full, non-Linblad master equation}

If one retains the counter-rotating terms in Eq.~(\ref{eq:CLMaster}), one finds corrections to the result Eq.~(\ref{eq:ThermDeph}).  These corrections are irrelevant in the small damping, small coupling ($\gamma, \lambda \ll 1$) limit of interest.  Explicitly, expanding $\Gamma_{\varphi,th}$ in $\gamma$ for a fixed $\lambda$, one finds that corrections
to Eq.~(\ref{eq:ThermDeph}) only arise at order $\gamma^3$:
\begin{eqnarray}
	\left[ \Gamma_{\varphi,th} \right]_{full} 
		& = & 
			 \gamma \nbar 
			 - 
			 \\
		&& \frac{\gamma^3}{4 \lambda^2} \frac{1 - 3 \lambda^2 + 4 \lambda^4}
			 {\left(1 - \lambda^2 \right)^2}
			 \nbar (1 + \nbar) (1 + 2 \nbar)
			\nonumber \\
			&& + O(\gamma^5) \nonumber
		 	\label{eq:ThermDephFull}
\end{eqnarray}
Comparing against the corresponding expansion of the approximate answer,  Eq.~(\ref{eq:ThermDeph2}), we see that the correction terms arising from using the full equation are suppressed by a factor $\lambda^2$ compared to the order-$\gamma^3$ terms in approximate answer..

Similarly, if we expand the full answer to lowest order $\lambda$, we find that
corrections to Eq.~(\ref{eq:ThermDephSmallLam}) are suppressed by $\gamma^2$:
\begin{eqnarray}
	\left[ \Gamma_{\varphi,th} \right]_{full} 
		& = & 
			\lambda^2 \frac{4 \nbar (1 + \nbar)}{\gamma} 
			\left(1 + \frac{\gamma^2}{4}  \right) + 
			O(\lambda^4) 
			\nonumber \\
		 	\label{eq:ThermDephFull2}
\end{eqnarray}
	
\subsection{Results for a counter-rotating qubit-oscillator interaction}

One can also calculate the thermal dephasing rate of the qubit using an qubit-oscillator interaction Hamiltonian which retains counter-rotating terms, c.f. Eq.~(\ref{eq:Hint2}).  It was recently suggested that such terms could play a significant role \cite{Serban06}.  Repeating the treatment which led to the full dephasing rate above (i.e.~no small damping approximation), we obtain:
\begin{eqnarray}
&&
	\left[ \Gamma_{\varphi,th} \right]_{H_{int,2}}  = 
	\nonumber \\
&&
	\frac{\gamma}{2} \textrm{Re } \left[
		A
		\sqrt{ \left(
				1 + \frac{2 i \lambda}{\gamma} 
			\right)^2
				+ \frac{8 i \lambda}{\gamma} \nbar
				- \frac{\gamma^2}{4}  }
		- 1
	\right]
\end{eqnarray}
where
\begin{eqnarray}
	A & = &
		\frac{2}{\sqrt{
		2 - \gamma^2 + 2 
			\sqrt{1 + 4 i \gamma \lambda (2 \nbar + 1) - 4 \lambda^2}
		}}
\end{eqnarray}				  
Note that this expression is identical to the previous Eq.~(\ref{eq:ThermDeph}) except for terms which are small as $\gamma$ or $\lambda$.  To see this explicitly, we can expand this result for small $\lambda$ to obtain:
\begin{eqnarray}
&&
	\left[ \Gamma_{\varphi,th} \right]_{H_{int,2}}  = 
	\nonumber \\
&&
		\frac{\lambda^2}{\gamma} \left[
			4 \nbar (1 + \nbar) 
			+ \gamma^2 \left(1 + 2 \nbar \right)^2
	\right]
	+ O(\lambda^4)
		\label{eq:ThermDephHint2Exp}
\end{eqnarray}
The second term above corresponds to the correction coming from using the non-rotating wave interaction Hamiltonian; it is suppressed by a factor $\gamma^2$.
Thus, as claimed, the non-rotating terms in the qubit-oscillator interaction Hamiltonian $H_{int}$ play no role in the weak damping, weak coupling ($\lambda \ll1, \gamma \ll \Omega$) limit of interest.  Note that $\Gamma_{\varphi,th}$ for an interaction Hamiltonain $H_{int,2}$ was also calculated in Ref.~\onlinecite{Serban06}; they found a result similar to Eq.~(\ref{eq:ThermDephHint2Exp}) above.

\section{Number statistics for a driven thermal mode}

In this appendix, we calculate the generating function for the number distribution for a driven harmonic oscillator which is also coupled to an equilibrium dissipative bath.  The Wigner function for such a state will be a Gaussian, with the variances determined by equipartition.  The means of the Gaussian will follow the equations of motion for a damped, driven oscillator.  In general, we can represent this state in the form:
\begin{eqnarray}
	\rho_{driven} = \hat{D} \rho_0 \hat{D}^{-1}
\end{eqnarray}
where $\rho_0$ is the equilibrium density matrix of the oscillator, and $\hat{D}$ is the displacement operator:
\begin{eqnarray}
	\hat{D} = \exp \left( \alpha_0 \hat{a}^{\dagger} - \alpha_0^* \hat{a} \right)
\end{eqnarray}
Finally, $\alpha_0$ is the coherent state amplitude corresponding to the driven, damped motion of the oscillator; only its magnitude will be important for the number statistics.  We have:
\begin{eqnarray}
	| \alpha_0 |^2 = \frac{ | F|^2 / 4 }{ (\omega_d - \Omega)^2 + \gamma^2 / 4 }
\end{eqnarray}

To proceed, it is useful to represent $\rho_0$ using a P function:
\begin{eqnarray}
	\rho_0 & = & \int d \alpha P_0(\alpha) | \alpha \rangle \langle \alpha | \\
	P_0(\alpha) & = & 
		\frac{1}{\pi} (e^{\beta \hbar \Omega} - 1) \exp \left( - |\alpha|^2 ( e^{\beta \Omega} - 1 ) \right)\\
\end{eqnarray}

It thus follows:
\begin{eqnarray}
	\rho & = &
		\int d \alpha P_0(\alpha) | \alpha + \alpha_0  \rangle \langle \alpha + \alpha_0 | 
\end{eqnarray}

The number distribution is then given by:
\begin{eqnarray}
	P(n) & = & \langle n | \rho  | n \rangle \\
	  & = &
		\int d \alpha P_0(\alpha) e^{-|\alpha+ \alpha_0|^2} \frac{ |\alpha+ \alpha_0|^{2n}}{n!} \\
	& = &
		\frac{ e^{\beta \Omega}-1 }{ \pi } 
		\int_0^{\infty} r dr \int_0^{2 \pi} d \theta
		\nonumber \\
		&&
			e^{-r^2 e^{\beta \Omega} - 2 r |\alpha_0| \cos \theta - |\alpha_0|^2}
			\frac{ \left( r^2 + |\alpha_0|^2 + 2 r \alpha_0 \cos \theta \right)^n }{n!}
			\nonumber \\
\end{eqnarray} 
	
The corresponding generating function is thus:
\begin{widetext}
\begin{eqnarray}
	P(k) & \equiv & \sum_n P(n) e^{i k n} \\
	& = &
		\exp \left[ - |\alpha_0|^2 (1 - e^{i k} ) \right] 
		\left(
			\frac{e^{\beta \omega}-1}{\pi}
		\right)
		\int_0^{\infty} r dr \int_0^{2 \pi} d \theta
		\exp\left[ -r^2 ( e^{\beta \omega}- e^{i k} ) \right] 
		\exp \left[ - 2 r |\alpha_0| \cos \theta (1 - e^{i k} ) \right]
		\nonumber \\
	& = &
		\exp \left[ - |\alpha_0|^2 (1 - e^{i k} ) \right] 
		\left[
		2 \left( e^{\beta \omega}-1\right)
		\int_0^{\infty} r dr 
 		\exp\left[ -r^2 ( e^{\beta \omega}- e^{i k} ) \right]
		I_0\left[ 2 r | \alpha_0 | (1 - e^{i k}) \right] \right]
		\nonumber
\end{eqnarray}
\end{widetext}
The first exponential factor above corresponds a generating function for a Poisson distribution; this would correspond to the number statistics of a driven, zero temperature oscillator (i.e.~coherent state statistics).  The second term corresponds to the effect of a non-zero temperature.  In the absence of the Bessel-function factor, it would yield the generating function for a thermal distribution. In the presence of a drive, the Bessel-function factors modifies the form of this term.  Thus, not surprisingly, the number statistics of a driven thermal oscillator is {\it not} simply a convolution of a Bose-Einstein distribution with a Poisson distribution.

We can evaluate the expression for $P(k)$ further by making use of:
\begin{eqnarray}
	I_0(z) = \sum_{l=0}^{\infty} \frac{z^{2 l}}{4^l (l!)^2} 
\end{eqnarray}
and
\begin{eqnarray}
	\int_0^{\infty} dr r^{1 + 2 m} e^{-A r^2} = \frac{m!}{2} A^{-m -1}
\end{eqnarray}

Some algebra thus yields:
\begin{eqnarray}
	P(k) & = & P_{therm}(k) \times P_{drive}(k) \times P_{corr}(k)
\end{eqnarray}
with
\begin{eqnarray}
	P_{therm} (k) & = & 
		\frac{1 - e^{-\beta \Omega}} {1 - e^{-\beta \Omega}e^{i k} } 
		\label{eq:GenFunc1} \\
	P_{drive} (k) & = & 
			\exp \left[ - |\alpha_0|^2 (1 - e^{i k} ) \right]  
			\label{eq:GenFunc2} \\
	P_{corr}(k) & = &
		\sum_{l=0}^{\infty} \frac{|\alpha_0|^{2l}  e^{- l \beta \Omega}}
			{ l! \left( 1 - e^{-\beta \Omega} e^{i k} \right)^l }
			\left( 1 - e^{i k}  \right)^{2l}		\nonumber \\
	& = &
		\exp\left[ |\alpha_0|^{2}  e^{-  \beta \Omega}
			 \frac{\left( 1 - e^{i k}  \right)^{2} }
			{  1 - e^{-\beta \Omega} e^{i k}  } \label{eq:GenFunc3}
		\right]
\end{eqnarray}

\end{appendix}

\bibliographystyle{apsrev}
\bibliography{AashRefs}

\end{document}